\documentclass{raa}

\usepackage{graphicx,times,xcolor}             
\usepackage{natbib}
\usepackage{amssymb,amsmath}
\bibpunct{(}{)}{;}{a}{}{,}

\usepackage[pagebackref=true]{hyperref}

\newcommand{\g}{$\gamma$-ray}
\newcommand{\lat}{\textit{Fermi}-LAT}
\newcommand{\obj}{V1405~Cas}

\begin{document}

  \title{Multi-wavelength study for gamma-ray nova V1405 Cas
}

   \author{Zi-wei Ou
      \inst{1,2}
   \and Pak-hin Thomas Tam
      \inst{2,3}
   \and Hui-hui Wang
      \inst{2,3}
   \and Song-peng Pei
      \inst{4}
   \and Wen-jun Huang
      \inst{2,3}
   }

   \institute{Tsung-Dao Lee Institute, Shanghai Jiao Tong University,
             Shanghai 201210, People's Republic of China; {\it ouziwei.astro@outlook.com}\\
        \and
             School of Physics and Astronomy, Sun Yat-Sen University, Zhuhai 519082, People's Republic of China; {\it tanbxuan@sysu.edu.cn}\\
        \and
            CSST Science Center for the Guangdong-Hongkong-Macau Greater Bay Area, Sun Yat-Sen University, Zhuhai 519082, People's Republic of China
        \and
             School of Physics and Electrical Engineering, Liupanshui Normal University, Liupanshui 553004, People's Republic of China\\
\vs\no
   {\small}}

\abstract{Novae are found to have GeV to TeV \g\ emission, which reveals the shock acceleration from the white dwarfs. Recently, \obj\ was reported to radiated suspicious \g\ by \textit{Fermi}-LAT with low significance ($4.1 \sigma$) after the optical maximum. Radio observations reveal that it is one of the five brightest novae surrounded by low-density ionized gas columns. Here we report continuous search for GeV \g\ from \textit{Fermi}-LAT. No \g\ were found. For \obj\, the flux level is lower than other well-studied \textit{Fermi} novae, and the \g\ maximum appear at $t_{0} + 145$ d. \g\ of \obj\ are used to search potential \g\ periodicity. No \g\ periodicity was found during the time of observation. By comparing multi-wavelength data, the \g\ upper limit to optical flux ratio with value at around $10^{-4}$ is obtained to constrain the shock acceleration. Long-term analysis from \textit{Swift}-XRT gets X-ray spectral in the post-shock phase, which indicates that \obj\ became a super-soft source. The best-fit black body temperature at the super soft state is 0.11 - 0.19 keV. 
}

   \authorrunning{Ou et al.}            
   \titlerunning{V1405 Cas}  
   \maketitle

\section{Introduction}         \label{sect:intro}

Classical nova is one kind of optical transient with brightness increase by one order in less than a few days \citep{Gallagher1978}. Such transient is an eruption that happens to a white dwarf (WD) which accretes from a main sequence star in close binary system \citep{DellaValle2020,Chomiuk2021a}. They are different from the so-called `luminous red novae' which are believed to arise from stellar mergers \citep{Pastorello2019,Cai2019,Cai2022Univ, Cai2022AA}. Classical novae are important objects for studying shock, nucleosynthesis and binary evolution \citep{Jose2006}. Since WDs undergoing nova eruptions can obtain mass during accretion, novae have been suggested as the progenitors of Type Ia supernovae in degenerate scenario \citep{Shara2010,Soraisam2015}. The shell of WD will be heated by the compression and undergo a thermonuclear runaway, which results in the ejection of the accreted mass \citep{Starrfield2016}. Such transient objects observed from radio to \g\ are identified firstly by optical observations. 

Since the \textit{Fermi} spacecraft launched in August 2008, more and more \g\ novae are discovered by the Large Area Telescope (LAT) on board \textit{Fermi} \citep{Ackermann2014,Franckowiak2018}. Nova can accelerate a fraction of the swept-up particles to high energies by diffusive shock acceleration within the \textit{Fermi}-LAT detection energy range. Both classical and recurrent novae (e.g. RS Oph) are detected by \textit{Fermi}-LAT \citep{Cheung2022}. RS Oph is seen even in TeV range also \citep{Acciari2022,Aharonian2022}. The GeV \g\ emission from Galactic novae observed by \textit{Fermi}-LAT reveal that these objects may accelerate relativistic particles by shocks probably \citep{Li2017}. The hadronic origin \g\ related to proton-proton collisions from novae imply a production of neutrinos \citep{Guepin2017,Fang2020}. For symbiotic novae (e.g. V407 Cyg, RS Oph), this emission is thought to have occurred when this high-velocity material shocked the dense stellar wind from their red giant companion. 

The ratio of \g\ and optical luminosity is suspected to be a key to constrain the lower limit on the fraction of the shock power that accelerate relativistic particles \citep{Metzger2015}. Therefore, simultaneous \g\ and optical observations would be necessary for Galactic novae. The typical \g\ novae are transient sources detected over 2 - 3 week duration \citep{Ackermann2014}. By assuming all novae are \g\ emitters, classical novae with $m_{R} \leq 12$ and within $\approx 8\,\rm kpc$ are likely to be discovered in \g\ using \textit{Fermi}-LAT \citep{Morris2017}.

Shock acceleration from nova not only emit \g\ but also radiate X-rays. The internal shocks from novae have velocities around $1000\,\rm km\, s^{-1}$ and heat the post-shock gas to temperatures above $10^{7}\,\rm K$, which can produce X-ray emission \citep{Steinberg2020}. Most \g\ novae show X-ray evidence of hot shock plasma, but not until the \g\ has faded below detectability \citep{Gordon2021}. Same as classical novae, the recurrent nova RS Oph were reported to produce X-ray emission originating from shocked ejecta before the X-ray supersoft source (SSS) emerged \citep{Orio2022b}. X-ray emission from classical novae during their \g\ period could be absorbed by dense ejecta \citep{Metzger2014}. 

\obj\ (PNV J23244760+6111140) was discovered on 18 March 2021 at 10:10 UT \citep{Wischnewski2022}.
Surprisingly, a series of optical flares were found with the brightest one reaching V=5.1 around 2021-05-10. Suspected \g\ emission has been found after 2021-05-10 with low detection significance, making it as one of the promising \g\ novae \citep{Gong2021,Buson2021}. However, the time difference between potential \g\ and optical is unclear. By using 4-day time bins, it was detected  with $4.1 \sigma$ significance in data from 2021-05-20 15:01:17 to 2021-05-24 15:01:17 UTC with a flux (E \textgreater 100 MeV) of $(1.4 \pm 0.8) \times 10^{-7}$ photons $\rm cm^{-2} s^{-1}$. Therefore, it is worthwhile to investigate \g\ emission for long term. Accordingly, one can check whether \obj\ has a \g\ emission occurring at late epoch compared to the other novae. Furthermore, no matter for significant flux or upper limit, shock acceleration of a nova may be constrained probably. In this paper, we investigate \g\ and X-ray emissions resulting from shock of \obj . The paper is organized in the following. Section \ref{sec:obs} gives data selection and analysis method. Section \ref{sec:result} shows the results from \g\ and X-ray data analysis. Section \ref{sec:discuss} discusses particle acceleration, white dwarf spinning and super soft X-ray state. Section \ref{subsec:conclu} summarizes our conclusions.

\section{Observations and Analysis} \label{sec:obs}

\subsection{AAVSO} \label{sec:aavso}

Optical light-curve of \obj\ has been collected from the American Association of Variable Star Observers (AAVSO)\footnote{http://www.aavso.org} \citep{Percy1993}. Coordinates of AAVSO observations for \obj\ give: R.A.=23$^h$24$^m$48$^s$, Dec.=61$^{\circ}$11'15" \citep{Wischnewski2022}. The optical maxima appears at MJD=59344.297. Several optical bands are included in data of AAVSO database. To put it simply, we use V band (Johnson V filter, Effective Wavelength = 5448$\,\rm \AA$) to generate light curve (see Fig. \ref{fig:mwl-lc} panel (a)). A galactic reddening of E(B-V) = 0.32 mag ($A_V$ = 1.03 mag with $R_V$ = 3.1) was assumed \citep{Schlafly2011}. $(B-V)_0$ of the earliest time (MJD 59293) and the peak time show  0.20 mag and 0.26 mag respectively, corresponding to colour temperatures $T_c$ of 7790 K and 7300 K \citep{Kitchin2013}. There is one significant peak in Episode I and several peaks in Episode II and III (definition of three episodes can be seen in Section \ref{sec:fermi-obs}).

The magnitude starts from 8.822 at the beginning and arise to 5.082 at MJD of 59344.3,
which is the first peak and the maximum of the whole time span. Then the magnitude go through fluctuation and arise to 5.91 as second peak, which locate in Episode II. After that, the magnitude continued to go through several times of slight fluctuation and then decline.

\subsection{\textit{Fermi}-LAT Observations} \label{sec:fermi-obs}

\textit{Fermi}-LAT photon data with energy range from 300 MeV to 100 GeV within 15$^{\circ}$ search radius are used in this analysis. Coordinates of $\textit{Fermi}$-LAT observations for \obj\ are: R.A.= 351.199$^{\circ}$, Dec.=61.1874$^{\circ}$ after running \texttt{gtfindsrc}. Events with a zenith angle greater than 90$^{\circ}$ were excluded. The Instrument Response Functions (IRFs) used in this analysis is P8R3\_SOURCE\_V3. The binned maximum-likelihood analysis (\texttt{gtlike}) was performed based on the 4FGL catalog (gll\_psc\_v27.fit). The Galactic emission gll\_iem\_v07 and isotropic diffuse emission iso\_P8R3\_SOURCE\_V3\_v1 are adopted. For sources within 5$^{\circ}$ from \obj\, the normalization parameters are freed. The spectrum of \obj\ is assumed to be a power law (PL):

\begin{equation}
    \frac{dN}{dE} = N_{0} (\frac{E}{E_{0}}) ^{\Gamma}
\end{equation}

where $N_{0}$ is prefactor in unit of $\rm cm^{-2}\,s^{-1}\,MeV^{-1}$, $\Gamma$ is the power index, $E_{0}$ is energy scale in unit of MeV.

To improve our starting model before analyzing the nova eruption, we fit a 1 yr data set spanning from 2020 March 18 to 2021 February 18, which ends 30 days before $t_{0}$. There is no significant detection at the position of \obj\ with a 95\% confidence flux upper limit, $\le 3.24 \times 10^{-9} \,\rm ph \,\rm cm^{-2} \,\rm s^{-1}$. As comparison, \g\ emission upper limits after $t_{0}$ are around $10^{-8} \,\rm ph \,\rm cm^{-2} \,\rm s^{-1}$, which would be shown in next section.

Upper limits at the 90\% confidence level (CL) are shown when the test statistic (TS) value is smaller than 4. The \textit{Fermi}-LAT light curve with 3-day bin from 2021-03-11 is shown. In addition, for comparing with the long-term X-ray observation, a 30-day bin analysis with PL is also performed.

At the same time, we assume an exponential cutoff power law (PLSuperExpCutoff) for \textit{Fermi}-LAT data to generate a light curve:

\begin{equation}
{dN \over dE} = N_0 \left( {E \over E_0}\right)^{\Gamma} exp\left( - \left({E \over E_c}\right)^{b} \right)
\end{equation}

where $N_{0}$ is prefactor in unit of $\rm cm^{-2}\,s^{-1}\,MeV^{-1}$, $\Gamma$ is power index, $E_{0}$ is energy scale in unit of MeV, $E_c$ is cutoff energy in unit of MeV, b is the second power index.

\subsection{\textit{Swift}-XRT Observations} \label{sec:swift-obs}

The Neil Gehrels Swift Observatory is a rapid-response satellite. The X-ray Telescope (XRT) on board is a focusing telescope which detect energy range between 0.2 and 10 keV. We selected \textit{Swift}-XRT observations of exposure time longer than 1 ks to check the evolution of the spectrum. An absorbed PL model was chosen to fit each observed spectrum based on web-form on the building. The information can be seen in Table \ref{tab:swift-log}.

The XRT cleaned event-lists were generated by the pipeline tool \texttt{xrtpipeline}. The source image, spectrum and light curve were extracted by \texttt{xselect}. Ancillary Response Files (ARFs) were created by using \texttt{xrtmkarf}. 

According to the \textit{Fermi}-LAT light curve, we divide \textit{Swift}-XRT observations into three episodes:
\begin{itemize}

\item[*] Episode I: Before MJD 59356, no \g\ are detected by \textit{Fermi}-LAT. The optical light curve rises to the maximum (the first peak).
\item[*] Episode II: From MJD 59356 to 59458, suspected \g\ were begun to be detected by \textit{Fermi}-LAT. The optical emissions arrive the second peak.  
\item[*] Episode III: After MJD 59458: the \g\ disappear.  

\end{itemize}

We combine the observations using \texttt{xselect}. For Episode I and Episode II, we can not conduct spectrum due to low counts rate. We only analyze Episode III.

\begin{figure*}
    \centering
    \includegraphics[scale=0.5]{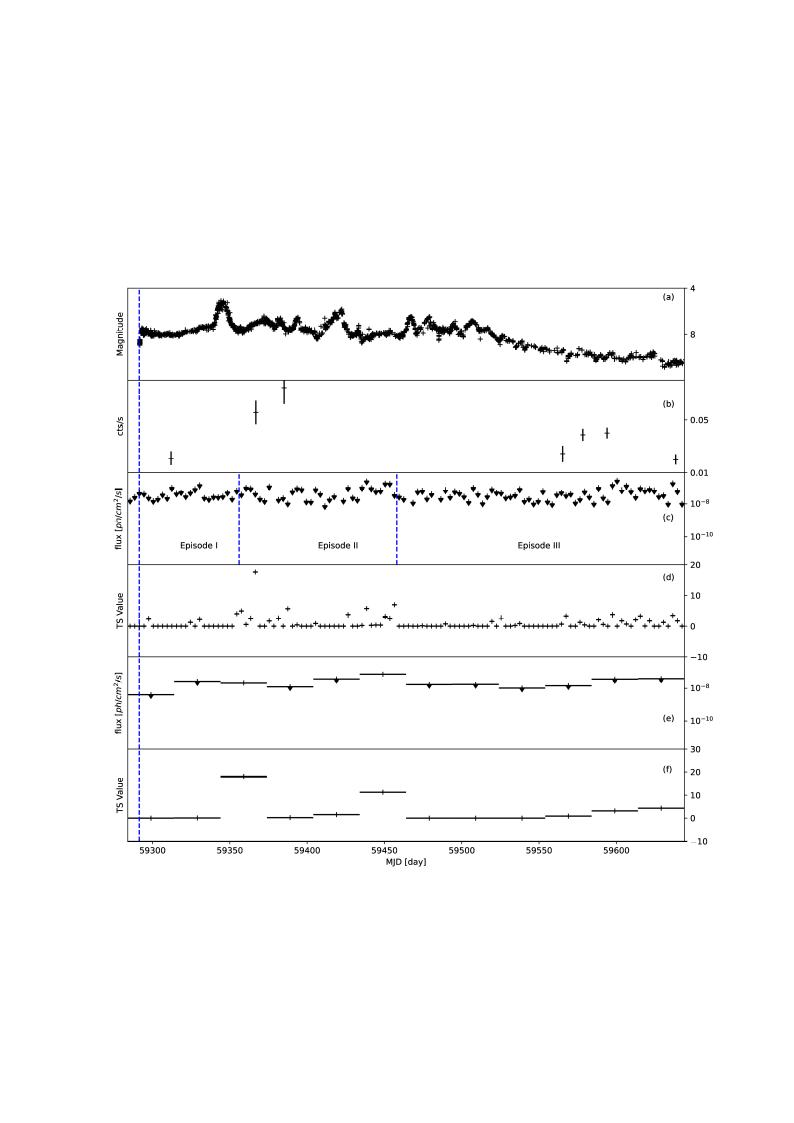}
    \caption{From top to bottom: (a) AAVSO V-band light curve; (b) \textit{Swift}-XRT light curve; (c) \textit{Fermi}-LAT 3-days bin upper limits (PL) with 90\% CL; (d) \textit{Fermi}-LAT 3-days TS value (PL); (e) \textit{Fermi}-LAT 30-days bin upper limits (PL) with 90\% CL; (f)\textit{Fermi}-LAT 30-days TS value (PL). The blue dashed lines divide the time into different episode. The red crosses indicate data points with TS values larger than 4.}
    \label{fig:mwl-lc}
\end{figure*}

\begin{figure}
    \centering
    \includegraphics[width=0.9\textwidth]{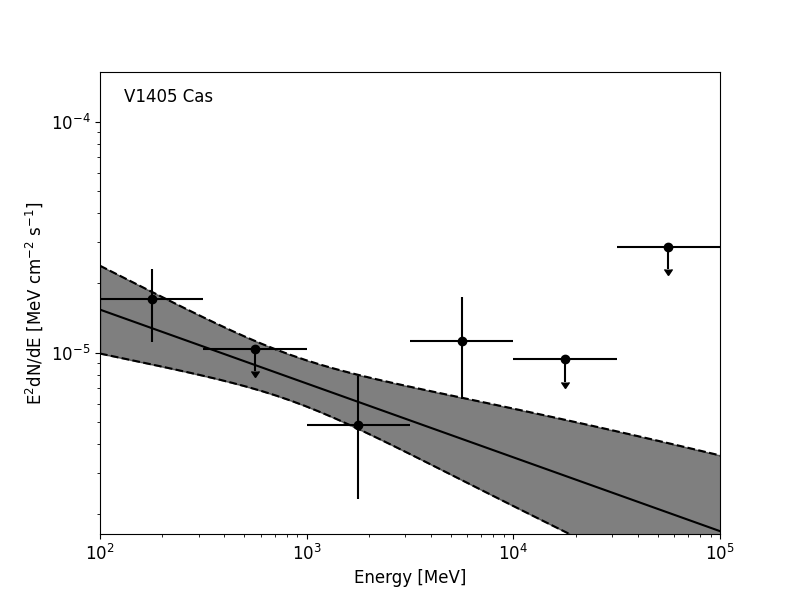}
    \caption{SED of $t_{0}$+65 $t_{0}$+77 with fixed $\Gamma$ to -2.3. The dashed lines and gray region indicate the $1 \sigma$ error of the model. The 95 \% CL upper limit are given also.}
    \label{fig:15days_sed}
\end{figure}

\section{Results} \label{sec:result}

\subsection{\textit{Fermi}-LAT light curve and spectrum}

From Fig. \ref{fig:mwl-lc}, we can see that TS $\geq 4$ \g\ appears around two weeks after the optical maximum. Since we use 3-day bins, we did not get exactly same results as \cite{Buson2021}. $t_{0}$ is defined as the discovery date (2021-03-18 10:10:00). For the PL model at time period of $t_{0} + 65$ d to $t_{0} + 68$ d and $t_{0} + 74$ d to $t_{0} + 77$ d, we get \g\ emission. We use the period from $t_{0} + 65$ d to $t_{0} + 77$ d to make a similar analysis (Fig. \ref{fig:15days_sed}), which gets the photon index of 1.94 and TS value of 26.9. A binned likelihood were performed based on \texttt{gtlike}. The corresponding best-fit parameters are shown in Table \ref{tab:fermi_fit}. Considering \obj\ as a point source, PL and PLSuperExpCutoff are compared. For PLSuperExpCutoff model, considering $E_c$ = 3000 MeV, the significance is slightly higher than those for $E_c$ = 3000 MeV. Since the number of free parameters are different, it is hard to draw a conclusion about the best fitting model.

\begin{figure}
    \centering
    \includegraphics[width=0.9\textwidth]{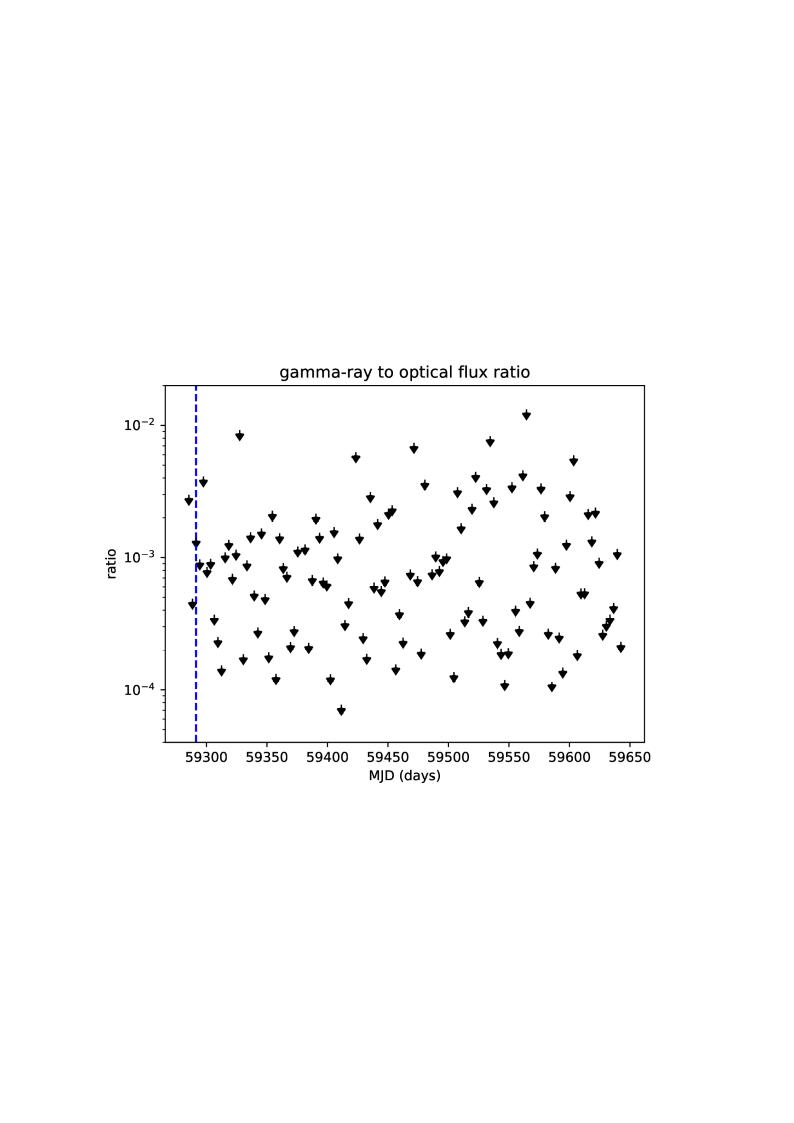}
    \caption{The ratio of \g\ upper limit to optical flux in Episode II. The blue dashed line gives the optical eruption time.}
    \label{fig:ratio}
\end{figure}

To investigate shock acceleration, we collect 3-days bin \g\ data at Figure \ref{fig:mwl-lc} to calculate the optical and \g\ flux ratio. The optical flux are calculated by averaging 3-days V band magnitude following $m/M = -2.5\, \log_{10} (f/F)$, where $m$ and $f$ are magnitude and brightness while $M$ and $F$ are absolute magnitude and absolute brightness. We average the optical flux error bar in each 3-day bin in order to match the \g\ upper limit and generate ratio error bar. The corresponding results are shown in Figure \ref{fig:ratio}. The occasional flux ratio is smaller than ASASSN-16ma value concluded by \cite{Li2017}, RS Oph value revealed by \cite{Cheung2022} and of other novae mentioned at Figure 9 of \cite{Chomiuk2021a}. This indicates that the internal shock is weaker than that in ASASSN-16ma.  

\begin{figure}
    \centering
    \includegraphics[width=0.9\textwidth]{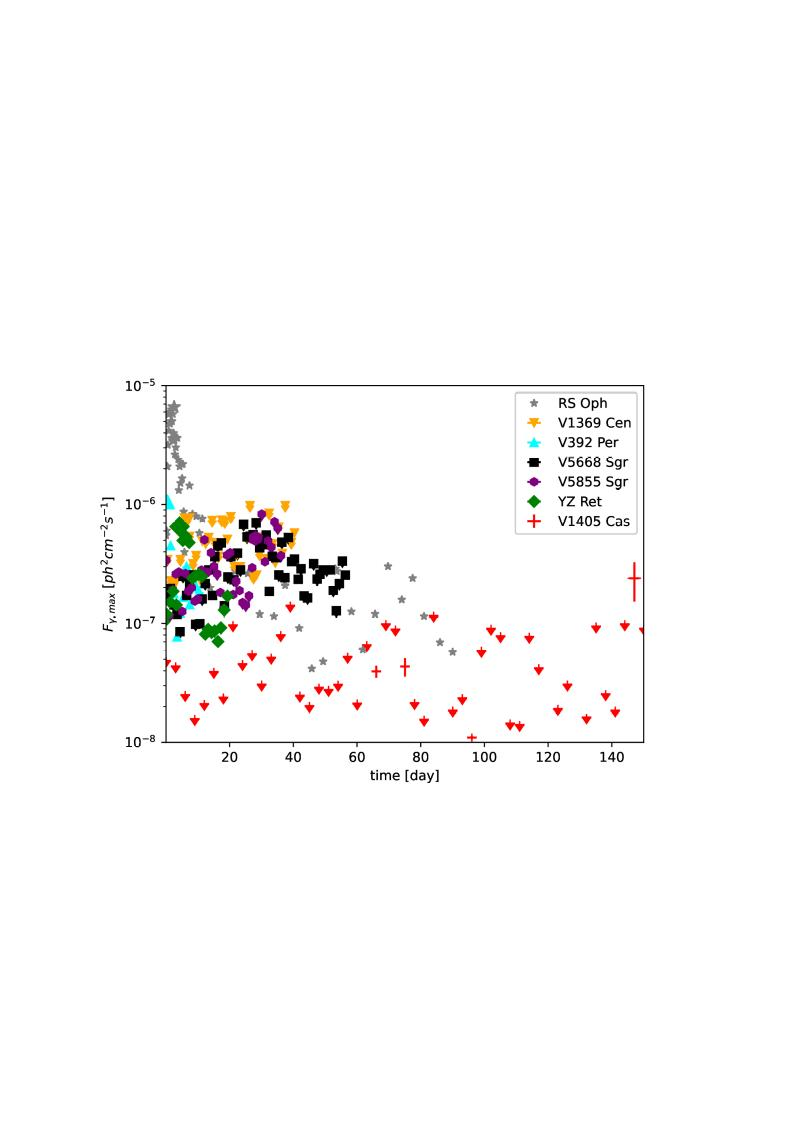}
    \caption{Well-studied \textit{Fermi}-LAT nova with known light curve. Reference: RS Oph \citep{Cheung2022}; V1369 Cen \citep{Cheung2016}; V392 Per \citep{Albert2022}; V5668 Sgr \citep{Cheung2016}; V5855 Sgr \citep{Nelson2019}; YZ Ret \citep{Sokolovsky2022}; \obj\ (this work). For X axis, Time [day] = 0 are corresponding to the optical eruption time for each nova. All the light curve are generated using 1-day bin, except to RS Oph and \obj\ . The light curve of RS Oph shows various time bin according to the orbit-timescale of 60 minutes, 6 hr, 1 day and 4 day. The light curve of \obj\ is binned into 3-day as mentioned previously.}
    \label{fig:all_nova}
\end{figure}

In Fig. \ref{fig:all_nova}, we compare light curve of well-studied \textit{Fermi} novae with \obj\ \footnote{https://asd.gsfc.nasa.gov/Koji.Mukai/novae/latnovae.html}. The flux upper limit of \obj\ is lower than all the other \textit{Fermi} novae. The $t = 0$ is defined as the optical maxima of novae. We can see that most \textit{Fermi} novae have \g\ maxima in one day to one month after the optical maxima.

\subsection{Gamma-ray periodicity} \label{subsec:1405period}

\begin{table}
  \caption{Best-fit parameters obtained by the analysis of \obj\ observation with \textit{Fermi}-LAT from $t_0$+65d to $t_0$+77d. Columns give (1) spectral model (2) flux, (3) power index, (4) cutoff energy, (5) second power index, (6) TS value.}{
  \begin{tabular}{llllll}
    \hline\hline
    spectral model & Flux ($\rm ph\,cm^{-2}\,s^{-1}$) & $\Gamma$ & $E_c$ (MeV) & b & TS \\\hline
    PL & $1.69\times10^{-8} \pm 7.88\times10^{-9}$ & $- 1.89\pm 0.29$ & ... & ... & 22.726\\
    & $2.50\times10^{-8} \pm 4.32\times10^{-9}$ & $- 2.3$ (fixed) & ... & ... & 22.293\\\hline
    PLSuperExpCutoff & $9.19\times10^{-9} \pm 3.16\times10^{-9}$ & $- 1.14\times10^{-5} \pm 0.62\times10^{-5}$ & 2000 (fixed) & 1 & 26.503\\ 
    & $9.92\times10^{-9} \pm 3.72\times10^{-9}$ & $- 0.44 \pm 0.37$ & 3000 (fixed) & 1 & 28.7199\\\hline
  \end{tabular}}\label{tab:fermi_fit}
\end{table}

Inspired by \cite{Li2022}, we search potential GeV \g\ periodicity of \obj\ . An aperture radius of $0.5^{\circ}$ was adopted for pulsation searching. We selected the \textit{Fermi}-LAT data at the time between $t_{0} + 65$ d and $t_{0} + 77$ d. The arrival time was corrected by \texttt{gtbary}. This tool performs a barycentering time correction to an event file using spacecraft orbit files of \lat\ . The selected events are analyzed by using \texttt{efsearch} task in the \texttt{HEASoft} package (version 6.31.1). \texttt{efsearch} is used to search for periodicity in a time series by folding the data over a range of periods. This search is motivated by the 544.84 s \g\ pulsation detected in nova ASASSN-16ma \citep{Li2022}. After performing a test run from 10 s to 800 s in a resolution of 0.1 s, we did not get any detection. In addition, Z-test is used to search periodicity also \citep{Kerr2011}. TS for each photon is weighted by its probability using the instrument response function. No periodicity are found. 

\subsection{X-ray spectral analysis}

For investigating the SSS phase, we consider an absorbed black body model for Swift-XRT observations in Episode III using online build Swift-XRT product \footnote{https://www.swift.ac.uk/user\_objects/}. Considering the SSS phase, the pile-up characteristics will differ from normal sources. This occurs when several photons hit the detector at the same place between different readouts. As a result, they are counted as one and their energies are summed. Both flux measurements and spectral characteristics would be affected by pipe-up. Selecting this method will cause only grade 0 (single pixel) events to be selected and a lower count-rate threshold to be used when correcting for pile-up. The corresponding best-fit parameters are shown in Table \ref{tab:bb_fit}, respectively. The column density of the $N_{\rm H}$ is $\sim 10^{21}\,\rm cm^{-2}$. The black body temperature kT distributes between 0.1 and 0.2 keV, which indicates that \obj\ became a SSS at the post-outburst state.

\begin{table}
    \caption{Best-fit parameters of black body model (0.3 - 6.7 keV). Columns give (1) observation ID, (2) column density of the neutral Hydrogen, (3) black body temperature, (4) unabsorbed Flux, (5) flux to count rate ratio, (6) fit W / degrees of freedom.}{%
    \begin{tabular}{cccccc}
        \hline\hline
        {ObsID} & $N_{\rm H}$  & {kT} & {$F_{un}$} & $\eta$  & {W/dof}\\ 
        & ($10^{21}\,\rm cm^{-2}$) & (keV) & ($\rm erg\, cm^{-2}\, s^{-1}$) & ($\rm erg\, cm^{-2}\, ct^{-1}$) & \\
        \hline
    00014197018 & $4^{+6}_{-3}$    & $0.13^{+0.06}_{-0.05}$ & $3.2^{+20.3}_{-2.4} \times 10^{-12}$    & $1.83 \times 10^{-11}$ & 33.57/27\\
    00014197020 & $1.4^{+2.4}_{-1.4}$  & $0.19^{+0.06}_{-0.05}$ & $1.26^{+2.09}_{-0.55} \times 10^{-12}$  & $1.81 \times 10^{-11}$ & 45.67/51\\
    00014197024 & $6.1^{+10.4}_{-6.1}$ & $0.11^{+0.16}_{-0.05}$ & $7^{+62}_{-7} \times 10^{-12}$          & $1.95 \times 10^{-11}$ & 16.3/26\\ \hline
    \end{tabular}}\label{tab:bb_fit}
\end{table}

\begin{table}
    \caption{Basic information of Swift-XRT observations. Columns give (1) observation ID, (2) exposure time, (3) observation start date, (4) time since the optical eruption.}{%
    \begin{tabular}{cccc}
        \hline\hline
        ObsID & Exposure (ks) & Date & $t_{0} +$ (d)\\
        \hline
         00014197001 & 0.28 & 2021-03-24T01:29:03 & 5.7\\ 
         00014197002 & 1.64 & 2021-03-28T18:10:40 & 10.4\\ 
         00014197003 & 1.85 &2021-03-31T08:19:03 & 12.9\\
         00014197004 & 1.57 & 2021-04-03T09:32:43 & 16.0\\
         00014197005 & 0.44 & 2021-04-06T18:47:32 & 19.4\\
         00014197006 & 1.83 & 2021-04-07T20:28:09 & 20.5\\
         00014197007 & 1.37 & 2021-04-09T04:14:48 & 21.8\\
         00014197008 & 1.42 & 2021-04-23T06:25:45 & 35.9\\
         00014197009 & 1.90 & 2021-04-30T02:12:33 & 42.7\\
         00014197010 & 1.80 & 2021-05-07T00:03:15 & 49.6\\
         00014197011 & 1.08 & 2021-05-14T16:44:50 & 57.3\\
         00014197012 & 0.92 & 2021-05-31T00:23:35 & 73.6\\
         00014197013 & 0.84 & 2021-06-15T07:23:29 & 88.9 \\
         00014197014 & 1.58 & 2021-06-30T11:45:17 & 104.1\\
         00014197015 & 0.73 & 2021-07-15T11:56:53 & 119.1\\
         00014197016 & 0.78 & 2021-07-30T04:06:01 & 133.8\\
         00014197017 & 1.04 & 2021-12-14T01:51:28 & 270.7\\
         00014197018 & 1.89 & 2021-12-15T00:03:15 & 271.6\\
         00014197019 & 1.29 & 2021-12-29T14:44:10 & 286.2\\
         00014197020 & 2.36 & 2022-01-14T09:35:17 & 302.0\\
         00014197024 & 2.34 & 2022-02-23T03:46:02 & 341.8\\ 
        \hline
    \end{tabular}}\label{tab:swift-log}
\end{table}

\section{Discussion} \label{sec:discuss}

\subsection{Particle acceleration and shock}

Figure \ref{fig:mwl-lc} shows the discrete \g\ upper limit concurrent with optical emissions (Episode II), which suggests that both are results of shock acceleration from nova explosion. GeV \g\ are thought to be the by-product of relativistic particles accelerated by shocks in the nova ejecta \citep{Figueira2018,Chomiuk2021a,Acciari2022}. Theoretically, the \g\ produced by novae are luminous, weighting in at $\sim$ 0.1 - 1\% of the bolometric luminosity predicts these events could generate photon energies up to 10 TeV, depending on details of the shocks \citep{Metzger2015}. This is verified by simultaneous \g\ and optical observations of Galactic nova ASASSN-16ma \citep{Li2017}.

According to Figure \ref{fig:mwl-lc}, the X-ray maximum is 20 days after the detection of the most significant \g\ emission. Thermal X-rays can be produced in the internal shocks that radiate the observed \g\ emission. One would expect to observe much brighter thermal X-ray emission at around 10 to 1000 times more luminous than the \g\ flux. In our case, simultaneous X-ray and \g\ observations of \obj\ at Episode II could constrain the radiative efficiency and the particle acceleration efficiency of the internal shocks. The presence of X-ray shock signatures sometimes coincident with the \g\ results from the small columns of neutral gas around this nova. 

Hybrid acceleration of leptons and hadrons in the nova shock has been considered in some research \citep{Martin2013}. In that model, the magnetic field  is obtained assuming an equipartition with the thermal energy density upstream of the shock. It results in the maximum energy of protons estimated to be $\sim$ 300 GeV. The GeV \g\ emission is then mostly expected from leptonic processes, namely the IC scattering of the nova light by the electrons accelerated in the shock. {However, hadronic and/or leptonic-hadronic models are favored to explain the gamma-ray emission of novae taking into account the energy loss of electrons via synchrotron emission \citep{Chomiuk2021a,DeSarkar2023}.} The narrow energy range at SED of \obj\ makes it hard to constrain hadronic or leptonic origins. 

\subsection{Whether \obj\ contains a spinning white dwarf?}

Unsurprisingly, a spinning white dwarf can produce gamma-rays like a spinning neutron star \citep{Takata2017,Orio2022a,Munari2022}. For example, the misalignment between axes of the WD spin, magnetic dipole, and orbital revolution produce \g\ pulsations. The nova progenitor is likely a fast-spinning magnetic WD, mostly an intermediate polar \cite{Orio1992}s. Strong bipolar nova winds from the magnetic poles of the WD regularly interact with the matter deposited into the orbital plane by the early and slow ejection \citep{Orio2022b}. Therefore, it is suitable to ask whether \obj\ are spinning, which accelerate electrons to produce GeV \g\ .

Previously, pulsations of nova are discovered on X-ray from V1674 Her and \g\ from ASASSN-16ma \citep{Drake2021, Li2022}. However, no \g\ pulsation were detected from \obj\ due to low photon counts. Nevertheless, this can not exclude \obj\ as a spinning WD (Section \ref{subsec:1405period}). More observations are needed to classify whether the WD \obj\ is spinning or not. 

\subsection{Super soft source} 

SSS is one kind of X-ray transient that is always believed to appear in post-outburst novae, in which hydrogen burning continues near the surface after the bulk of accreted envelope mass has been ejected \citep{Orio2022a}. From Table \ref{tab:bb_fit} we can see \obj\ is a SSS characterized by soft X-ray emissions. Such super-soft phase appear after shock emergence. The spectra in some novae in the supersoft X-ray phase may have redshift velocity of the emission lines and blueshift velocity of the absorption lines \citep{Pei2021}. The flux variation in some emission lines may result from the surrounding cool inhomogeneous material or temporary ionization. 

Radio observations from VLA and uGMRT reveal that \obj\ are among the five brightest novae ever observed in the radio band, and the expansion velocity to be 1200 km/s \citep{Chomiuk2021b,Nayana2022}. Such high velocity can heat the gas, which produces X-ray emission at the post-shock phase. The appearance of radio emissions when the \g\ detected by \textit{Fermi}-LAT results from the low-density columns of ionized gas. As expected, \obj\ has begun to display Fe II absorption features and then a SSS \citep{Shore2021,Page2021}. Thus, it becomes an excellent target to study SSS after shock acceleration. This Fe II absorption point to the nova formation in a large circumbinary envelope of gas \citep{Williams2012,Aydi2024}. The gas may have a origin of the secondary star.

The typical range of effective temperature $T_{\rm eff}$ in the SSS varies from $\sim 1.2 \times 10^5$ K to about $10^6$ K \citep{Suleimanov2003} . The low emission during the SSS phase can be explained by absorption from neutral and/or ionized medium \citep{Ness2023}. \obj\ has a typical temperature of SSS. For example, RS Oph has a blackbody temperature in the 35 to 40 eV ($\approx$ 405000 to 470000 K) range \citep{Orio2023}. V2491 Cyg has a temperature of 90 $\pm$ 1 eV ($10^6$ K) and an X-ray luminosity of $178 \times 10^{36}$ erg s$^{-1}$ \citep{Ness2022}. The temperature is proportional to the shock velocity. Typical soft X-ray photons are absorbed in the ISM with values of $N_{\rm H}$ in several $10^{21}\,\rm cm^{-2}$. As X-rays suffer photoelectric absorption, $N_{\rm H}$ reflect these effects, which is also the case for novae \citep{Nelson2019}. Possibly, the X-ray and \g\ originated from different region \citep{Metzger2015}. The $N_{\rm H}$ values of \obj\ are around $1.0\times 10^{21}\,\rm cm^{-2}$ to $1.6\times 10^{22}\,\rm cm^{-2}$. However, resulting from the low quality of X-ray data, we can not derive evolution of $N_{\rm H}$.

\section{Conclusion} \label{subsec:conclu}

We summarize our conclusion as follows. (1) Typical \g\ novae show significant \g\ after the optical maxima in weeks to months. Our result shows that no significant GeV \g\ emissions of nova \obj\ were detected before and after the optical maxima. (2) Typical \g\ novae have \g\ maxima at hours to weeks after optical maxima. Our result reveal that \g\ upper limit of nova \obj\ show a maximum at $t_0 + 145$ d. (3) Some classical novae were found to have periodicity in \g\ emission. We search \g\ periodicity in \obj\ using two methods: \texttt{efsearch} and Z-test. Owing to low GeV photon counts, no \g\ periodicity is found from \textit{Fermi}-LAT. (4) X-ray emission of novae arrives at a soft phase. Our work introduce an adsorb black body model to investigate the potential SSS phase. We show that \obj\ turned to a super-soft state after \g\ emissions disappeared according to the best-fit temperature values. (5) \g\ to optical ratio is a key for shock acceleration, our result shows a ratio around $10^{-4}$ to $10^{-2}$ which approach bright \g\ novae (e.g., ASASSN-16ma, RS Oph).

Although some novae have low significance and weak signal in \g\ , it is still valuable to observe such objects. With higher sensitivity, the future Very Large Area Gamma-ray Space Telescope (VLAST) may improve that \citep{Fan2022}. X-ray telescope, such as Einstein Probe (EP) and Space-based multi-band astronomical Variable Objects Monitor (SVOM), would also help us to unveil SSS nature for more nova.

\begin{acknowledgements}

The authors thanks Kwan-Lok Li for many useful discussion. Scientific results from data presented in this publication are obtained from AAVSO and HEASARC. Z.W.O. is supported by the National Natural Science Foundation of China (NSFC, Grant No. 12393853). P.H.T., H.H.W. and W.J.H. are supported by the National Natural Science Foundation of China (NSFC) under grant 12273122. H.H.W. is supported by the Scientific Research Foundation of Hunan Provincial Education Department (21C0343). S.P.P. is supported by Science Research Project of University (Youth Project) of the department of education of Guizhou Province (QJJ[2022]348) and the Science and Technology Foundation of Guizhou Province (QKHJC-ZK[2023]442).

\end{acknowledgements}

\bibliographystyle{raa}
\bibliography{bibtex}

\begin{thebibliography}{57}
\providecommand\natexlab[1]{#1}
\providecommand\JournalTitle[1]{#1}

\bibitem[{Acciari} {et~al.}(2022)]{Acciari2022}
{Acciari}, V.~A., {Ansoldi}, S., {Antonelli}, L.~A., {et~al.} 2022, Nature Astronomy, 6, 689

\bibitem[{Ackermann} {et~al.}(2014)]{Ackermann2014}
{Ackermann}, M., {Ajello}, M., {Albert}, A., {et~al.} 2014, Science, 345, 554

\bibitem[{Albert} {et~al.}(2022)]{Albert2022}
{Albert}, A., {Alfaro}, R., {Alvarez}, C., {et~al.} 2022, \apj, 940, 141

\bibitem[{Aydi} {et~al.}(2024)]{Aydi2024}
{Aydi}, E., {Chomiuk}, L., {Strader}, J., {et~al.} 2024, \mnras, 527, 9303

\bibitem[{Buson} {et~al.}(2021)]{Buson2021}
{Buson}, S., {Cheung}, C.~C., \& {Jean}, P. 2021, The Astronomer's Telegram, 14658, 1

\bibitem[{Cai} {et~al.}(2022{\natexlab{a}})]{Cai2022Univ}
{Cai}, Y., {Reguitti}, A., {Valerin}, G., \& {Wang}, X. 2022{\natexlab{a}}, Universe, 8, 493

\bibitem[{Cai} {et~al.}(2019)]{Cai2019}
{Cai}, Y.~Z., {Pastorello}, A., {Fraser}, M., {et~al.} 2019, \aap, 632, L6

\bibitem[{Cai} {et~al.}(2022{\natexlab{b}})]{Cai2022AA}
{Cai}, Y.~Z., {Pastorello}, A., {Fraser}, M., {et~al.} 2022{\natexlab{b}}, \aap, 667, A4

\bibitem[{Cheung} {et~al.}(2016)]{Cheung2016}
{Cheung}, C.~C., {Jean}, P., {Shore}, S.~N., {et~al.} 2016, \apj, 826, 142

\bibitem[{Cheung} {et~al.}(2022)]{Cheung2022}
{Cheung}, C.~C., {Johnson}, T.~J., {Jean}, P., {et~al.} 2022, \apj, 935, 44

\bibitem[{Chomiuk} {et~al.}(2021{\natexlab{a}})]{Chomiuk2021a}
{Chomiuk}, L., {Metzger}, B.~D., \& {Shen}, K.~J. 2021{\natexlab{a}}, \araa, 59, 391

\bibitem[{Chomiuk} {et~al.}(2021{\natexlab{b}})]{Chomiuk2021b}
{Chomiuk}, L., {Linford}, J.~D., {Aydi}, E., {et~al.} 2021{\natexlab{b}}, \apjs, 257, 49

\bibitem[{De Sarkar} {et~al.}(2023)]{DeSarkar2023}
{De Sarkar}, A., {Nayana}, A.~J., {Roy}, N., {Razzaque}, S., \& {Anupama}, G.~C. 2023, \apj, 951, 62

\bibitem[{Della Valle} \& {Izzo}(2020)]{DellaValle2020}
{Della Valle}, M., \& {Izzo}, L. 2020, \aapr, 28, 3

\bibitem[{Drake} {et~al.}(2021)]{Drake2021}
{Drake}, J.~J., {Ness}, J.-U., {Page}, K.~L., {et~al.} 2021, \apjl, 922, L42

\bibitem[{Fan} {et~al.}(2022)]{Fan2022}
{Fan}, Y.~Z., {Chang}, J., {Guo}, J.~H., {et~al.} 2022, Acta Astronomica Sinica, 63, 27

\bibitem[{Fang} {et~al.}(2020)]{Fang2020}
{Fang}, K., {Metzger}, B.~D., {Vurm}, I., {Aydi}, E., \& {Chomiuk}, L. 2020, \apj, 904, 4

\bibitem[{Figueira} {et~al.}(2018)]{Figueira2018}
{Figueira}, J., {Jos{\'e}}, J., {Garc{\'\i}a-Berro}, E., {et~al.} 2018, \aap, 613, A8

\bibitem[{Franckowiak} {et~al.}(2018)]{Franckowiak2018}
{Franckowiak}, A., {Jean}, P., {Wood}, M., {Cheung}, C.~C., \& {Buson}, S. 2018, \aap, 609, A120

\bibitem[{Gallagher} \& {Starrfield}(1978)]{Gallagher1978}
{Gallagher}, J.~S., \& {Starrfield}, S. 1978, \araa, 16, 171

\bibitem[{Gong} \& {Li}(2021)]{Gong2021}
{Gong}, Y.-H., \& {Li}, K.-L. 2021, The Astronomer's Telegram, 14620, 1

\bibitem[{Gordon} {et~al.}(2021)]{Gordon2021}
{Gordon}, A.~C., {Aydi}, E., {Page}, K.~L., {et~al.} 2021, \apj, 910, 134

\bibitem[{Gu{\'e}pin} \& {Kotera}(2017)]{Guepin2017}
{Gu{\'e}pin}, C., \& {Kotera}, K. 2017, \aap, 603, A76

\bibitem[{H.~E.~S.~S. Collaboration} {et~al.}(2022)]{Aharonian2022}
{H.~E.~S.~S. Collaboration}, {Aharonian}, F., {Ait Benkhali}, F., {et~al.} 2022, Science, 376, 77

\bibitem[{Jos{\'e}} {et~al.}(2006)]{Jose2006}
{Jos{\'e}}, J., {Hernanz}, M., \& {Iliadis}, C. 2006, \nphysa, 777, 550

\bibitem[{Kerr}(2011)]{Kerr2011}
{Kerr}, M. 2011, \apj, 732, 38

\bibitem[{Kitchin}(2013)]{Kitchin2013}
{Kitchin}, C.~R. 2013, {Astrophysical Techniques}

\bibitem[{Li}(2022)]{Li2022}
{Li}, K.-L. 2022, \apjl, 924, L17

\bibitem[{Li} {et~al.}(2017)]{Li2017}
{Li}, K.-L., {Metzger}, B.~D., {Chomiuk}, L., {et~al.} 2017, Nature Astronomy, 1, 697

\bibitem[{Martin} \& {Dubus}(2013)]{Martin2013}
{Martin}, P., \& {Dubus}, G. 2013, \aap, 551, A37

\bibitem[{Metzger} {et~al.}(2015)]{Metzger2015}
{Metzger}, B.~D., {Finzell}, T., {Vurm}, I., {et~al.} 2015, \mnras, 450, 2739

\bibitem[{Metzger} {et~al.}(2014)]{Metzger2014}
{Metzger}, B.~D., {Hasco{\"e}t}, R., {Vurm}, I., {et~al.} 2014, \mnras, 442, 713

\bibitem[{Morris} {et~al.}(2017)]{Morris2017}
{Morris}, P.~J., {Cotter}, G., {Brown}, A.~M., \& {Chadwick}, P.~M. 2017, \mnras, 465, 1218

\bibitem[{Munari} {et~al.}(2022)]{Munari2022}
{Munari}, U., {Giroletti}, M., {Marcote}, B., {et~al.} 2022, \aap, 666, L6

\bibitem[{Nayana} {et~al.}(2022)]{Nayana2022}
{Nayana}, A.~J., {Anupama}, G.~C., {Banerjee}, D., {et~al.} 2022, The Astronomer's Telegram, 15383, 1

\bibitem[{Nelson} {et~al.}(2019)]{Nelson2019}
{Nelson}, T., {Mukai}, K., {Li}, K.-L., {et~al.} 2019, \apj, 872, 86

\bibitem[{Ness} {et~al.}(2022)]{Ness2022}
{Ness}, J.~U., {Beardmore}, A.~P., {Bezak}, P., {et~al.} 2022, \aap, 658, A169

\bibitem[{Ness} {et~al.}(2023)]{Ness2023}
{Ness}, J.~U., {Beardmore}, A.~P., {Bode}, M.~F., {et~al.} 2023, \aap, 670, A131

\bibitem[{Orio} {et~al.}(1992)]{Orio1992}
{Orio}, M., {Trussoni}, E., \& {Oegelman}, H. 1992, \aap, 257, 548

\bibitem[{Orio} {et~al.}(2022{\natexlab{a}})]{Orio2022a}
{Orio}, M., {Gendreau}, K., {Giese}, M., {et~al.} 2022{\natexlab{a}}, \apj, 932, 45

\bibitem[{Orio} {et~al.}(2022{\natexlab{b}})]{Orio2022b}
{Orio}, M., {Behar}, E., {Luna}, G.~J.~M., {et~al.} 2022{\natexlab{b}}, \apj, 938, 34

\bibitem[{Orio} {et~al.}(2023)]{Orio2023}
{Orio}, M., {Gendreau}, K., {Giese}, M., {et~al.} 2023, \apj, 955, 37

\bibitem[{Page} {et~al.}(2021)]{Page2021}
{Page}, K.~L., {Starrfield}, S., {Munari}, U., {Woodward}, C.~E., \& {Wagner}, R.~M. 2021, The Astronomer's Telegram, 15111, 1

\bibitem[{Pastorello} {et~al.}(2019)]{Pastorello2019}
{Pastorello}, A., {Mason}, E., {Taubenberger}, S., {et~al.} 2019, \aap, 630, A75

\bibitem[{Pei} {et~al.}(2021)]{Pei2021}
{Pei}, S., {Orio}, M., {Ness}, J.-U., \& {Ospina}, N. 2021, \mnras, 507, 2073

\bibitem[{Percy} \& {Mattei}(1993)]{Percy1993}
{Percy}, J.~R., \& {Mattei}, J.~A. 1993, \apss, 210, 137

\bibitem[{Schlafly} \& {Finkbeiner}(2011)]{Schlafly2011}
{Schlafly}, E.~F., \& {Finkbeiner}, D.~P. 2011, \apj, 737, 103

\bibitem[{Shara} {et~al.}(2010)]{Shara2010}
{Shara}, M.~M., {Yaron}, O., {Prialnik}, D., \& {Kovetz}, A. 2010, \apjl, 712, L143

\bibitem[{Shore} {et~al.}(2021)]{Shore2021}
{Shore}, S.~N., {Buil}, C., {Dubovsky}, P., {et~al.} 2021, The Astronomer's Telegram, 14577, 1

\bibitem[{Sokolovsky} {et~al.}(2022)]{Sokolovsky2022}
{Sokolovsky}, K.~V., {Li}, K.-L., {Lopes de Oliveira}, R., {et~al.} 2022, \mnras, 514, 2239

\bibitem[{Soraisam} \& {Gilfanov}(2015)]{Soraisam2015}
{Soraisam}, M.~D., \& {Gilfanov}, M. 2015, \aap, 583, A140

\bibitem[{Starrfield} {et~al.}(2016)]{Starrfield2016}
{Starrfield}, S., {Iliadis}, C., \& {Hix}, W.~R. 2016, \pasp, 128, 051001

\bibitem[{Steinberg} \& {Metzger}(2020)]{Steinberg2020}
{Steinberg}, E., \& {Metzger}, B.~D. 2020, \mnras, 491, 4232

\bibitem[{Suleimanov} {et~al.}(2003)]{Suleimanov2003}
{Suleimanov}, V., {Meyer}, F., \& {Meyer-Hofmeister}, E. 2003, \aap, 401, 1009

\bibitem[{Takata} {et~al.}(2017)]{Takata2017}
{Takata}, J., {Yang}, H., \& {Cheng}, K.~S. 2017, \apj, 851, 143

\bibitem[{Williams}(2012)]{Williams2012}
{Williams}, R. 2012, \aj, 144, 98

\bibitem[{Wischnewski}(2022)]{Wischnewski2022}
{Wischnewski}, E. 2022, BAV Magazine Spectroscopy, 11, 6

\end{thebibliography}

\label{lastpage}

\end{document}